\documentclass[10pt,conference]{IEEEtran}
\usepackage{amsfonts, amsmath, amssymb, epsfig, subfigure, graphicx, color}

\newtheorem{theorem}{Theorem}

\begin{document}

\title{Structured interference-mitigation in two-hop networks}

\author{\authorblockN{Yiwei Song}
\authorblockA{Department of Electrical and Computer Eng.\\
University of Illinois at Chicago\\
Chicago, IL, USA \\
Email: ysong34@uic.edu}
\and
\authorblockN{Natasha Devroye}
\authorblockA{Department of Electrical and Computer Eng.\\
University of Illinois at Chicago\\
Chicago, IL, USA \\
Email: devroye@uic.edu}}
%

\maketitle

\newcommand{\nrd}[1]{\textcolor{red}{#1}}
\newcommand{\nbl}[1]{\textcolor{blue}{#1}}

%
%

\begin{abstract}
We consider two-hop S-R-D Gaussian networks with a source (S), a relay (R) and a destination (D), some of which experience additive interference. This additive interference, which renders the channels state-dependent, is either a) experienced at the destination D and known non-causally at the source S, or b) experienced at the relay R and known at the destination D.  In both cases, one would hope to exploit this knowledge of the channel state at some of the nodes to obtain  ``clean'' or interference-free channels, just as Costa's dirty-paper coding does 
for one-hop channels with state non-causally known to the transmitter. We demonstrate a scheme which achieves to within $\frac{1}{2}$ bit of a  ``clean'' channel. This novel scheme is based on nested-lattice code and a Decode-and-Forward (DF) relay. Intuitively, this strategy uses the structure provided by nested lattice codes to cancel the ``integer'' (or lattice quantized) part of the interference and treats the ``residual'' (or quantization noise) as noise. 
\end{abstract}

\section{Introduction and channel model}

\subsection{Motivation}
\begin{figure}[h]
\centering
\includegraphics[width=9cm]{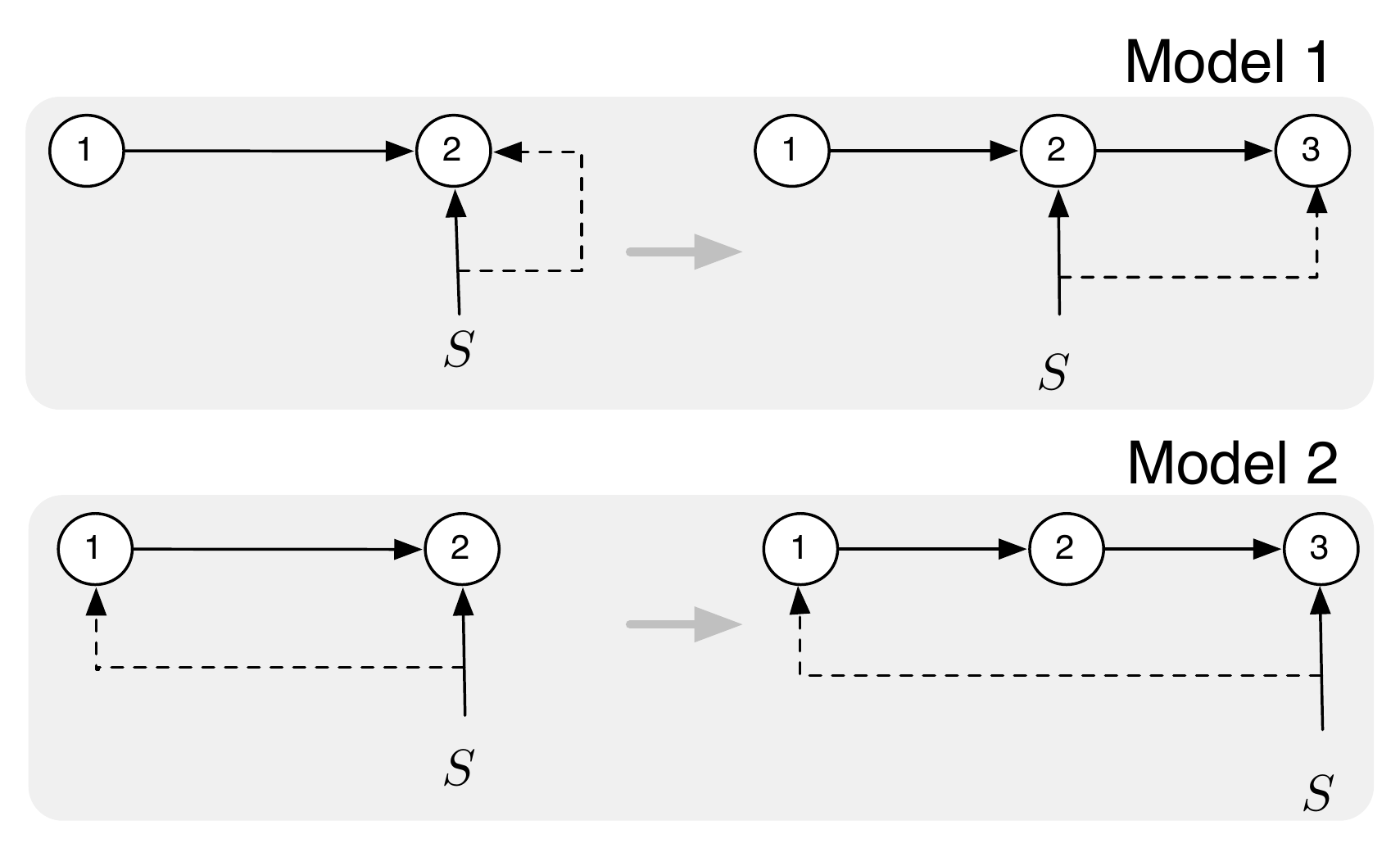}
\vspace{-0.5cm}
\caption{Left side: two channels with state information $S$ experienced at the nodes with a solid line, and known at the nodes with a dotted line. Right side: the two two-hop relay networks considered here. In Model 1, interference $S$ is experienced at the relay and known at the destination, while in Model 2, interference $S$ is experienced at the destination but only known at the source.}
\label{fig:model}
\end{figure}

In this paper, we examine the capacity of two-hop Gaussian relay networks with channel state information. This channel state, which in our channel models amounts of additive interference $S$ (which we emphasize may be arbitrary), is experienced at certain nodes, and  known at other nodes. While channels with state information have been considered extensively in the past, most work has considered a single-hop model, for example the channel with state information known non-causally at the encoder studied by Ge'fand and Pinsker \cite{gelfand} for general discrete memoryless channels and by Costa \cite{Costa:1983:DPC} for the Gaussian noise channel model (shown on the bottom right of Fig. \ref{fig:model}). Here, we consider two two-hop networks, as shown on the right-hand side of Fig. \ref{fig:model}, where solid lines indicated where $S$ is ``experienced'', while dotted lines indicate where that interference is known. In Model 1, the additive interference $S$ is experienced at relay but is known at the destination. In Model 2, the additive interference $S$ is experienced at the destination but is known at the transmitter.  In Fig. \ref{fig:model} the two two-hop channels considered here are plotted next to single-hop equivalents: channel with state information known at the receiver (Rx) and transmitter (Tx) respectively. The two-hope versions considered in Model 1 and Model 2 extend these more classical models through the introduction of a relay node. 
 
The question we seek to answer is whether, as in Costa's famous ``dirty-paper coding'' for a point-to-point channel with state known non-causally to the encoder, coding schemes may be derived that achieve the capacity of a two-hop {\it interference-free} channel. We note that this is not immediately obvious, as in our channel models, the Costa-like setup is not present, i.e. we do not consider a channel model where the relay experiences $S$, which is known at the transmitter. In this case, a direct application of Costa's dirty-paper coding on the first link would result in an interference-free two-hop channel. In short, we will demonstrate lattice-coding based strategies which achieve at most $\frac{1}{2}$ bit from the interference-free outer bound, with rates whose expressions are independent of $S$. 

Such two-hop networks are not only of theoretical interest, as they extend more classical ``channels with side-information'' of channels with state know at the encoders/decoders, to multi-hop scenarios, but they may also be motivated in networks with cognition, or where certain nodes have side-information about the messages or interference experienced at other nodes. How this information is obtained is beyond the scope of this work, but we suggest a possible motivation for channel Model 1:  consider the multi-hop / line network in which two messages are being transmitted: one from Node 1 to 3 through Node 2 (relay) and another from Node 3 to Node 4 further down the line. If Nodes 1 and 3 transmit simultaneously, the relay Node 2 may see interference from Node 3; but this ``interference'' is known at Node 3 in the next hop when it attempts to decode Node 1's message. In this scenario, simple dirty-paper coding techniques may not be immediately employed at the relay due to the presence of the potentially large amount of interference seen in decoding Node 1's message (which in turn is needed for a DPC re-encoding) at the relay. Thus, alternative schemes which in some way allow the interference to be forwarded and canceled by the receiver are needed. 
We will use structured codes -- nested-lattice codes in the additive white Gaussian noise (AWGN) channel models considered.

\subsection{Past work}
This work considers the capacity of channels with state known non-causally at some encoders/decoders. Such channels have been considered by Gel'fand and Pinsker \cite{gelfand}, whose result was applied to the AWGN channel by Costa \cite{Costa:1983:DPC} using the celebrated ``dirty-paper coding'' technique. A nice survey of other channels with state information is provided in \cite{keshet2007channel}. 
We will use nested-lattice codes in deriving achievable rate regions, as such codes have been shown to be capacity-achieving for the point-to-point AWGN channel \cite{Erez:2004}, to be capacity-approaching for two-way relay channels \cite{Wilson:2010, Nam:IEEE, Song:2010:latticerelay}, to be useful in the compute-and-foward framework \cite{Nazer:2009:computeandforward}, and finally to be ``good for almost everything'', as expanded upon in \cite{Erez:latticegood:2005}. 
The utility of lattice codes \cite{zamir-lattices} in achieving the capacity of certain classes of channels with side-information is considered in \cite{Zamir:2002:binning}.  The nested lattice approach of  \cite{Zamir:2002:binning} for the dirty-paper channel is extended to dirty-paper networks in \cite{Kim:2004:muDPC, Akhbari:2009:DirtyRelay, Zaidi:2010:DirtyRelay, Philosof:DirtyMAC}, where in some scenarios lattice codes are interestingly shown to outperform random codes. 
The most similar work to the considered two-hop relay network Model 1 is the work of 
\cite{Akhbari:2009:DirtyRelay} which considers a relay channel with state non-causally available at the {\it transmitter, relay, or both,}  and derives CF-based achievable rates for discrete memoryless channels. In \cite{Zaidi:2010:DirtyRelay} a three terminal relay channel is again considered where the state is known {\it only at the source,} where upper and lower bounds are derived in Gaussian noise. Our channel model differs in that 1)  we consider a two-hop network, and there is no direct link between transmitters and receivers as in the relay channel, and 2) the state is only known at the receiver rather than the transmitter and/or relay. Finally, a Compress-and-Forward (CF)-based scheme (rather than the DF-based schemes considered here) is proposed for Model 1 in the authors' submitted work \cite{Song:2011:CISS}.

\subsection{Contributions}
The central contribution of this work is the proposal of two novel lattice-code-based Decode-and-Forward (DF) schemes for the two-hop networks with side-information considered in Models 1 and 2 of Fig. \ref{fig:model}. These two schemes result in the identical achievable rates given in Theorem \ref{thm:main}, and utilize the structure of nested lattice codes to cancel the ``integer'' (or lattice quantized) part of the interference $S$ and treat the remaining ``residual'' (or quantization error) as noise. 



\section{Notation, nested lattice coding, and channel model preliminaries}
\label{sec:notation}
We briefly outline definitions and notation for nested lattice codes for transmission over AWGN channels, following those of \cite{zamir2002nested, nam:2009nested, Song:2010:latticerelay};  \cite{loeliger1997averaging, zamir2002nested,  Erez:2004} and in particular \cite{zamir-lattices} offer more thorough treatments, followed by more formal channel model definitions.

\subsection{Nested lattice codes}
\label{subsec:nested}

 An $n$-dimensional lattice $\Lambda$ is a discrete subgroup of Euclidean space $\mathbb{R}^n$ (of vectors ${\bf x}$, though we will denote these without the bold font as $x$) with Euclidean norm $|| \cdot ||$ under vector addition. We may define

$\bullet$ The {\it nearest neighbor lattice quantizer} of $\Lambda$ as \[ Q_\Lambda({x}) = \arg \min_{\lambda\in \Lambda} ||{x}-\lambda||;\]

$\bullet$ The { \texttt{mod }$\Lambda$} operation as ${x}$  \texttt{mod }$\Lambda : = {x} - Q_\Lambda({x})$, hence 
\[ {x} = Q_{\Lambda}({x}) + ({x} \mod \Lambda); \]

$\bullet$ The {\it fundamental region of $\Lambda$} as the set of all points closer to the origin than to any other lattice point \[\mathcal{V}(\Lambda):= \{{x}:Q({x}) = {\bf 0}\}\]
which is of volume $V: = \mbox{Vol}({\mathcal V}(\Lambda))$.

$\bullet$ The {\it second moment per dimension of a uniform distribution over ${\mathcal V}$} as
\[ \sigma^2(\Lambda) : = \frac{1}{V}\cdot \frac{1}{n} \int_{\mathcal V} ||{x}||^2 \; d{x}\]

$\bullet$ The {\it crypto lemma} \cite{Forney:ShannonWinener:2003}: which states that  $({\bf  x} + {\bf U}) \mod \Lambda$ (where ${\bf U}$ is uniformly distributed over $\mathcal{V}$) is an independent random variable uniformly distributed over $\mathcal{V}$.

Standard definitions of {\it Poltyrev good} and {\it Rogers good} lattices are used \cite{Erez:2004, Erez:latticegood:2005}, and by \cite{Erez:latticegood:2005} we are assured the existence of lattices which are both Polytrev and Rogers good, which may intuitively be thought of as being good channel and source codes, respectively. 

The proposed schemes will be based on {\it nested lattice codes}. To define these, consider two lattices $\Lambda$ and $\Lambda_c$ such that $\Lambda \subseteq \Lambda_c$ with fundamental regions ${\cal V}, {\cal V}_c$ of volumes $V, V_c$ respectively. 
Here $\Lambda$ is called the {\it coarse} lattice which is a sublattice of  $\Lambda_c$,  the {\it fine} lattice.  The set  $ \mathcal{C}_{\Lambda_c, {\cal V}} = \{ \Lambda_c \cap \mathcal{V} \} $ may be employed as the codebook for transmission over the AWGN channel, with coding rate $R$ defined as
\[ R = \frac{1}{n} \log |\mathcal{C}_{\Lambda_c, {\cal V}}| = \frac{1}{n} \log \frac{V}{V_c}.\] Here $\rho = |\mathcal{C}_{\Lambda_c, {\cal V}}|^{\frac{1}{n}} = \left( \frac{V}{V_c} \right)^{\frac{1}{n}}$ is the nesting ratio of  this {\it nested  $(\Lambda, \Lambda_c)$ lattice code}. A pair of {\it good} Nested lattice codes, where $\Lambda$ is both {\it Rogers good} and {\it Poltyrev good}  and $\Lambda_c$ is {\it Poltyrev good}, were shown to exist and be capacity achieving (as $n\rightarrow \infty$) for the AWGN channel \cite{Erez:2004}.

The {\it goodness} of lattice codes pair can be extended to a lattice codes chain. All the lattice codes in the nested lattice chain $\Lambda \subseteq \Lambda_1 \subseteq \Lambda_2$ can be both {\it Rogers good} and {\it Poltyrev good} \cite{Krithivasan:2007:goodlattice} for the arbitrary nesting ratios. {\it Good} lattice chains will be used in the achievability schemes for both models, described next. 

 \subsection{Channel models}
 \label{subsec:models}
 We consider two AWGN two-hop channel models with interference, as shown on the right hand side of Fig. \ref{fig:model}. In particular,  both Model 1 and Model 2 consist of three nodes, the ``source'' or Tx, node 1; the ``relay'' node 2, and the ``destination'' or Rx, node 3. The channel inputs of nodes 1 and 2 are denoted by $X_1$ and $X_2$, taking on values $x_1\in {\cal X}_1$ and $x_2\in {\cal X}_2$ subject to average power constraints $E[|X_1|^2]\leq P_1$ and $E[|X_2|^2]\leq P_2$. The received signals at node 2 and 3 respectively are $Y_2$ and $Y_3$, taking on values $y_2\in {\cal Y}_2$ and $y_3\in {\cal Y}_3$. We will communicate over $n$ channel uses and we let $X_1^n : = (X_1(1), X_1(2), \cdots X_1(n))$ where $X_1(k)$ denotes the input at channel use $k$ (and similarly for $X_2^n, Y_2^n, Y_3^n$. We consider two-hop AWGN networks with arbitrary interference $S^n$, where, at channel use $k$, the inputs and outputs of the channels are related as
 \begin{align*}
 \mbox{Model 1:} \;\; Y_2(k)& = X_1(k)+S(k)+Z_2(k), \\
 Y_3(k)&= X_2(k)+Z_3(k), \;  \mbox{ Rx 3 knows }S(k)\\
 \mbox{Model 1:} \;\; Y_2(k)& = X_1(k)+Z_2(k), \mbox{ Tx 1 knows }S(k) \\
 Y_3(k)& = X_2(k)+S(k)+Z_3(k),
 \end{align*} 
 where for notational convenience, it is assumed that power constraints of the source and relay are $1$, i.e. $P_1=P_2=1$, and the noise is AWGN with $Z_2(k)\sim {\cal N}(0,\frac{1}{S_1})$ and $Z_3(k)\sim {\cal N}(0,\frac{1}{S_2})$ respectively. This ensures that, in the absence of interference $S$, the link $1\rightarrow 2$ has capacity   $\frac{1}{2}\log(1+S_1)$ and the link $2\rightarrow 3$ has capacity $\frac{1}{2}\log(1+S_2)$. In this channel, we wish to transmit a message $w\in \{1,2,\cdots, 2^{nR}\}$ at rate $R>0$ from node 1 to node 3 (which forms as estimate $\hat{w}$ of $w$ from its received signal $Y_3^n$) such that $\Pr\{\hat{w}\neq w\}\rightarrow 0$ as the number of channel uses, $n\rightarrow \infty$. From now on, to simplify notation, we will abuse notation slightly and drop the superscript $n$, using $X_1$ to denote $X_1^n$ for the remainder, as we will always be dealing with $n$ channel uses. 
 
\section{Achievable rate for Models 1 and 2}
\label{sec:achievable}
Our main result is presented in Theorem \ref{thm:main} where we show two different achievability schemes for Models 1 and 2 which achieve the same rate.

\begin{theorem}
The following rate may be achieved using a structured nested-lattice coding based DF scheme for both Models 1 and 2:
\begin{align}
 R &< \left[ \frac{1}{2} \log \left( \frac{1}{ \frac{1}{1+S_1} + \frac{1}{1+ S_2}} \right) \right]^+  \\
 & =  \left[ \frac{1}{2} \log \left( \frac{S_1S_2 + S_1 + S_2+1}{S_1 + S_2 + 2} \right) \right]^+, 
 \end{align}
where $S_1$ and $S_2$ are the signal-to-noise ratio for the two links: $1 \rightarrow 2 $ and $2 \rightarrow 3$ respectively. 
For the special case $S_1 = S_2 = S$, this reduces to $R < \frac{1}{2} \log \left( \frac{1}{2} + \frac{S}{2} \right)$. 
\label{thm:main}
\end{theorem}

\smallskip

\noindent {\it Remark:} We note that the rate achieved in Theorem \ref{thm:main} achieves to within at most $\frac{1}{2}$ bit of  the clean channel capacity which forms an outer bound for both channel Models 1 and 2, as
\begin{align*}
\frac{1}{2} \log \left( \frac{1}{ \frac{1}{1+S_1} + \frac{1}{1+ S_2}} \right) + \frac{1}{2} &= \frac{1}{2} \log \left( \frac{1}{ \frac{1}{2(1+S_1)} + \frac{1}{2(1+ S_2)}} \right) \\
&\geq \frac{1}{2} \log \left( \frac{1}{ 2*\frac{1}{2(1+\min(S_1, S_2) ) } } \right) \\
&=\frac{1}{2} \log \left( 1 + \min(S_1, S_2) \right).
\end{align*}

We now prove Theorem \ref{thm:main} for both Model 1 and Model 2 in the following two subsections. For both achievability proofs, we consider a good nested lattice chain  $ \Lambda \subseteq \Lambda_c \subseteq \Lambda_q$ as in \ref{subsec:nested}, where
we will specify the second moment / power constraints in the following.  We need $\Lambda$ and $\Lambda_q$ to be both {\it Rogers good} and {\it Poltyrev good} , and $\Lambda_c$ to be {\it Poltyrev good}. The existence of such a good lattice chain is proved in \cite{Krithivasan:2007:goodlattice}.   The message coding rate is 
\[ R = \frac{1}{n} \log \left( \frac{V(\Lambda)}{V(\Lambda_c) } \right) = \frac{1}{2} \log \left( \frac{1}{\sigma^2(\Lambda_c) } \right) \]
The coding rate of $ Q_{\Lambda_q} ( ) \mod \Lambda $ is 
\[ R_q = \frac{1}{n} \log \left( \frac{V(\Lambda)}{V(\Lambda_q) } \right) = \frac{1}{2} \log \left( \frac{1}{\sigma^2(\Lambda_q) } \right).\]

\subsection{Achievability proof for Model 1}

\begin{figure*}[ht]
\centering
\includegraphics[width=16cm]{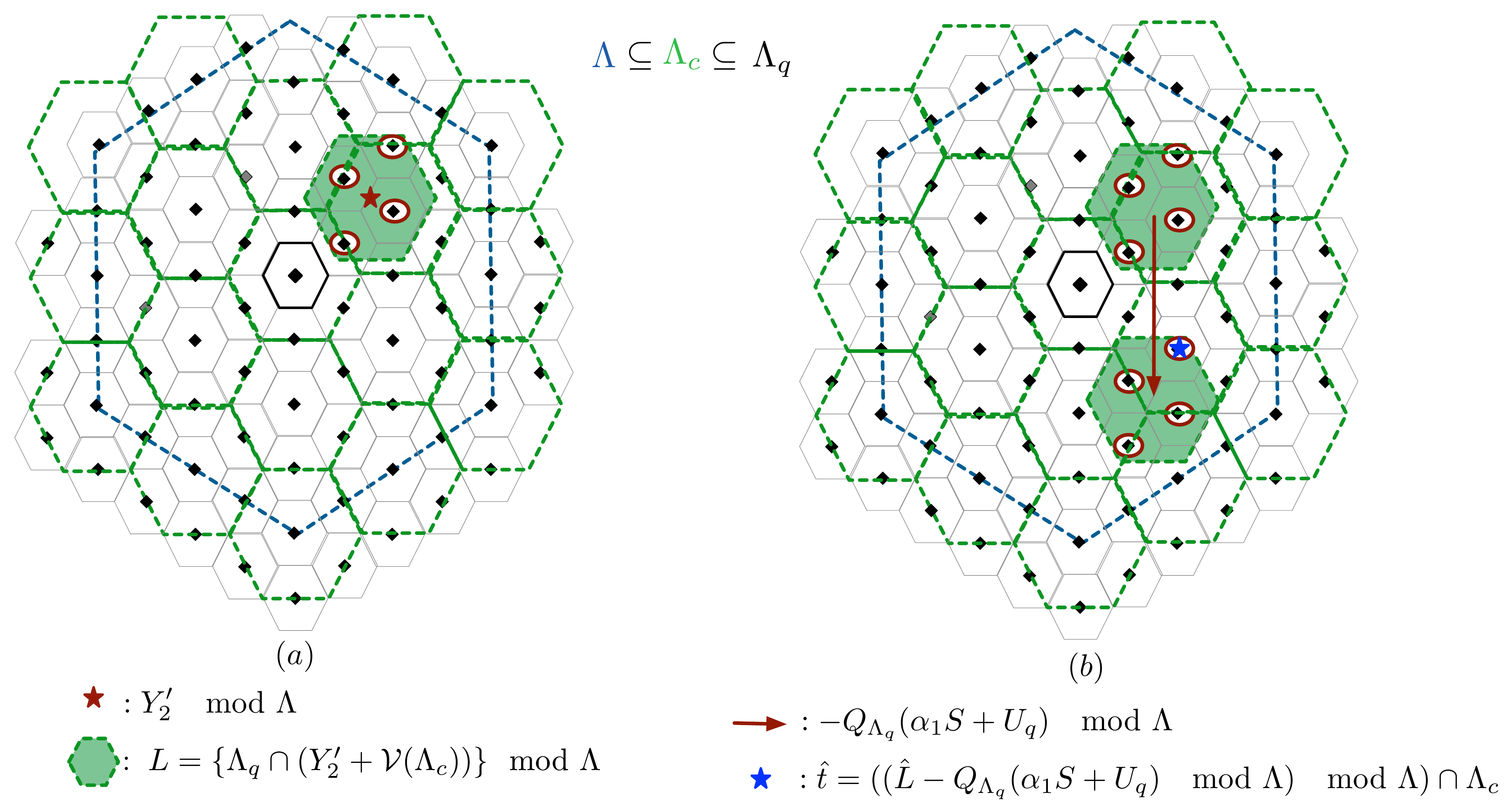}
\caption{Illustration of key steps of achievability scheme for Model 1: (a) the list decoding performed at the relay node 2, and (b) decoding and interference cancelation performed at the destination node 3.}
\label{fig:model1}
\end{figure*}

{\bf Encoding at the source (Node 1)} : message $w \in \mathcal{W} = \{ 1 ,2,  \dots, 2^{nR} \}$ is one-to-one mapped to the lattice codeword $t \in \{ \Lambda_c \cap \mathcal{V} (\Lambda) \} $ ($ w \leftrightarrow t  $). 
 The transmitter chooses the $t$ associated with transmitted message and sends 
\[ X_1 = ( t + U_1 ) \mod \Lambda \] 
 where $U_1$ is the dither uniformly distributed over $\mathcal{V}(\Lambda)$, which is known by relay. The second moment of $\Lambda$ is $\sigma^2(\Lambda) = P_1 = 1$. Also notice that $X_1$ is uniformly distributed over $\mathcal{V}(\Lambda)$ and independent of $t$.  

{\bf Decoding at the relay (Node 2)}: the relay receives 
\[ Y_2 = X_1 + S + Z_2 \]
and uses a Minimum Mean Squared Error (MMSE) estimator to decode $t$, by computing
\begin{align*}
Y_2' &= ( \alpha_1 Y_2 + U_q  - U_1 ) \mod \Lambda \\
&= ( \alpha_1 X_1 + \alpha_1 S + \alpha_1 Z_2 + U_q - U_1) \mod \Lambda \\
&= ( t + U_1 - (1- \alpha_1) X_1 + \alpha_1 S  + \alpha_1 Z_2+ U_q - U_1) \mod \Lambda \\
&= ( t + Q_{\Lambda_q} (\alpha_1 S + U_q)  + (\alpha_1 S + U_q) \mod \Lambda_q \\
& \;\;\;\;\;\;\;  - (1- \alpha_1) X_1 + \alpha_1 Z_2 ) \mod \Lambda  \\
\end{align*}
where $U_q$ is the quantization dither uniformly distributed over $\Lambda_q$, where we recall that
$\Lambda \subseteq \Lambda_c \subseteq \Lambda_q$. Thus, $ (\alpha_1 S + U_q) \mod \Lambda_q$, $- (1- \alpha_1) X_1$, and $\alpha_1 Z_2$ can be seen as three independent noise terms with variances $\sigma^2 (\Lambda_q)$, $(1 - \alpha_1)^2 $ and $\alpha_1^2 \frac{1}{S_1}$. These three terms approximate Gaussian noise as in \cite{Erez:2004, Song:2010:latticerelay} when $n \rightarrow \infty$.

Choosing $\alpha_1= \alpha_{1opt} = \frac{S_1}{S_1 + 1} $, the relay uses the nested lattice list decoding scheme of \cite{Song:2010:latticerelay} to decode a list  $L$ of terms of the form $ (t + Q_{\Lambda_q} (\alpha_1 S + U_q) )\mod \Lambda$. This step is illustrated in (a) of Fig. \ref{fig:model1}, and produces the list:
\begin{align*}
L = \{ \Lambda_{q} \cap (Y_2' + \mathcal{V}(\Lambda_c)) \} \mod \Lambda.
 \end{align*}
This list is guaranteed to have one codeword as the list decoding region is ${\cal V}(\Lambda_c)$. To ensure that the probability of error (or the probability that the correct codeword is in the list), as $n\rightarrow \infty$ we require,  
 treating two of the interference terms as noise, 
\begin{align}
R < \frac{1}{2} \log \left( \frac{1}{ \frac{1}{1+S_1} +\sigma^2(\Lambda_q)} \right). \label{x}
\end{align}
The number  of lattice codewords $(t - Q_{\Lambda_q}(\cdot) )\mod \Lambda$ in the list is $ 2^{n(R_q - R)}$, where $R_q = \frac{1}{2} \log(\frac{1}{\sigma^2(\Lambda_q)})$ is the coding rate of $Q_{\Lambda_q} ( ) \mod \Lambda $. The number of lists is equal to the number of finer lattice codewords, i.e.  $2^{nR_q}$. Notice that this step does not constrain the rate $R_q$, but that $R_q$ is always greater than $R$ since
\begin{align*}
R_q &= \frac{1}{2} \log \left( \frac{1}{\sigma^2(\Lambda_q)} \right) > \frac{1}{2} \log \left( \frac{1}{ \frac{1}{1+S_1} +\sigma^2(\Lambda_q)} \right)> R.
\end{align*}

{\bf Encoding at the relay and decoding at the destination}: The relay transmits the index of the list with any capacity achieving code (which may, but need not be a nested lattice code). The destination can decode the list index if
\begin{align*}
 R_q = \frac{1}{2} \log \left( \frac{1}{\sigma^2(\Lambda_q)} \right) < \frac{1}{2} \log ( 1 + S_2),
\end{align*}
constraining the second moment of the quantization lattice $\Lambda_q$  
\begin{align}
 \sigma^2(\Lambda_q) > \frac{1}{1 + S_2}. \label{z}
\end{align}
After decoding the index of the list $\hat{L}$, the destination can determine the transmitted codeword uniquely as 
\[ \hat{t} = (  ( \hat{L} - Q_{\Lambda_q} (\alpha_1 S + U_q) \mod \Lambda ) \mod \Lambda )  \; \cap \; \Lambda_c. \]
This last step is illustrated in Fig. \ref{fig:model1}(b).
Combining \eqref{x} and \eqref{z}, the achievable rate of the proposed scheme is 
\begin{align}
 R &< \frac{1}{2} \log \left( \frac{1}{\frac{1}{1+S_1} + \sigma^2(\Lambda_q) } \right) \notag  \\
 &<  \frac{1}{2} \log \left( \frac{1}{\frac{1}{1+S_1} +  \frac{1}{1 + S_2}   }\right)  \label{f}  \\
&=  \frac{1}{2} \log \left( \frac{S_1S_2 + S_1 + S_2+1}{S_1 + S_2 + 2} \right). \notag
\end{align}

\subsection{Achievability proof for Model 2}

\begin{figure*}[ht]
\centering
\includegraphics[width=16cm]{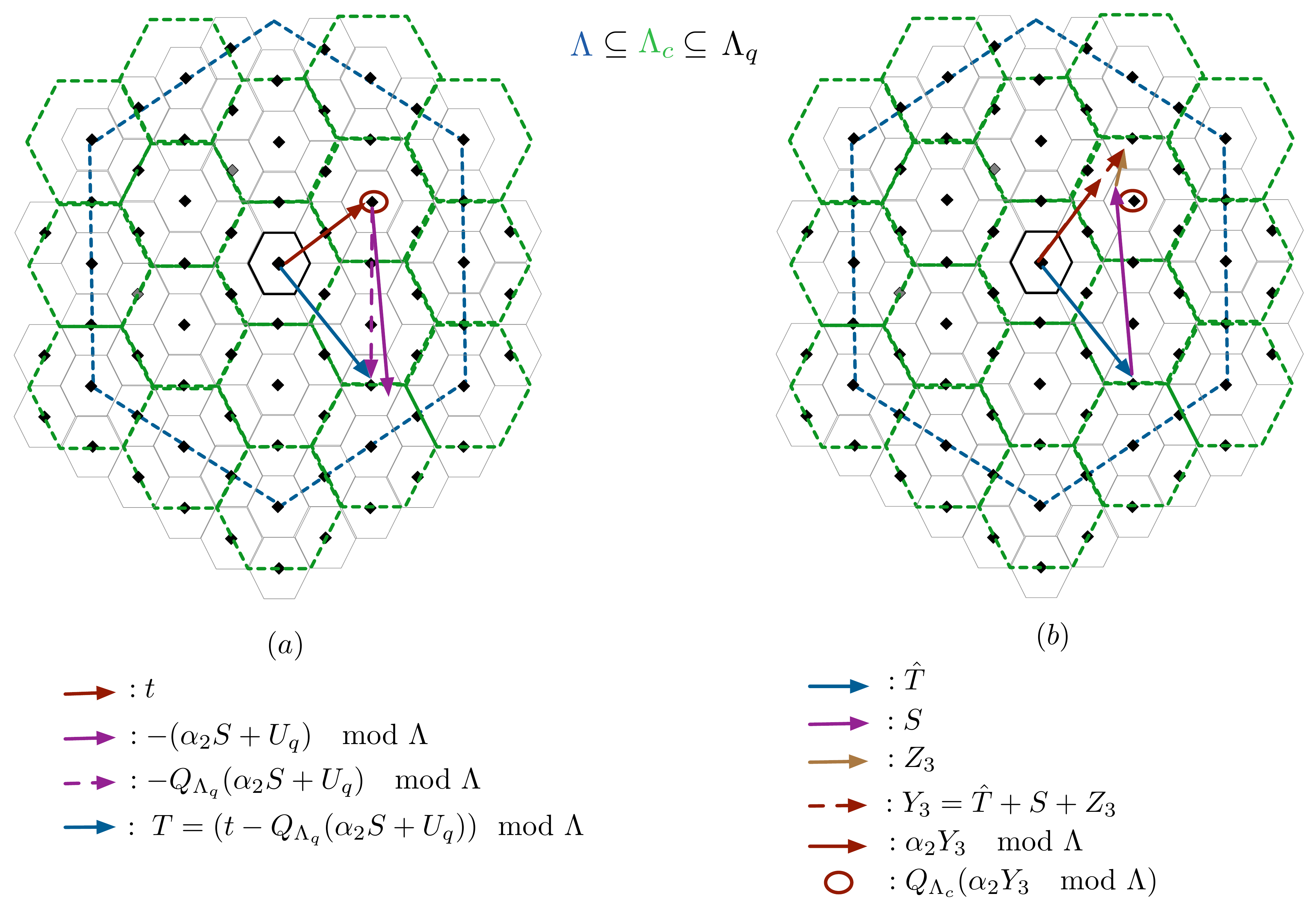}
\label{fig:model2}
\caption{Illustration of key achievability steps in the scheme used for Model 2: (a) interference ``pre-cancellation'' is performed at the transmitter Node 1 who knows the interference $S$, (b) the receiver Node 3 experiences interference $S$, and suffers from a quantization noise ``residual''.}
\end{figure*}

{\bf Encoding at the source (Node 1):}  message $w \in \mathcal{W} = \{ 1 ,2,  \dots, 2^{nR} \}$ is one-to-one mapped to the lattice codeword $t \in \{ \Lambda_c \cap \mathcal{V} (\Lambda) \} $ ($w \leftrightarrow t $).  
 The transmitter chooses the $t$ associated with transmitted message and sends 
\[ X_1 = ( T + U_1 ) \mod \Lambda \]
where $ T =  ( t - Q_{\Lambda_q}  (\alpha_2 S + U_q) ) \mod \Lambda$.
$U_q$ is the quantization dither which is uniformly distributed over $\mathcal{V} (\Lambda_q)$ and also known by the destination. Here $U_1$ is the channel coding dither which is uniformly distributed over $\mathcal{V} (\Lambda) $ and is known at the relay.  The second moment of $\Lambda$ is limited by the transmit power, which is assumed to be 1 in this case. This encoding step is illustrated in Fig \ref{fig:model2}(a), where we see the pre-subtraction of the scaled and quantized interference $S$, all $\mod \Lambda$.

{\bf Decoding at the relay (Node 2):}  the relay receives 
\[ Y_2 = X_1 + Z_2 \]
and forms the following signal
\begin{align*}
Y_2' &=  (\alpha_1 Y_2 - U_1) \mod \Lambda\\
&=  ( \alpha_1 X_1 + \alpha_1 Z_2 - U_1 ) \mod \Lambda \\ 
&= ( t - Q_{\Lambda_q} (\alpha_2 S + U_q) - (1-\alpha_1) X_1 + \alpha_1 Z_2 )\mod \Lambda.
\end{align*}
Choosing $\alpha_1 = \alpha_{1opt} = \frac{S_1}{S_1 + 1} $, the relay can decode $\widehat{T}$, an estimate of $T$ subject to the constraints:
\begin{align}
 R &< \frac{1}{2} \log ( 1 + S_1 ), \label{con1} \\
 R_q &< \frac{1}{2} \log ( 1 + S_1 ).\end{align}
The probability of error analysis is similar to \cite{Erez:2004, Nam:IEEE, Wilson:2010, Nazer:2009:computeandforward}. This implies that
\begin{align*}
 R_q = \frac{1}{2} \log \left( \frac{1}{ \sigma^2(\Lambda_q) }\right) < \frac{1}{2} \log ( 1 + S_1 )
\end{align*}
which results again in a lower bound on the quantization lattice's second moment
\[ \sigma^2(\Lambda_q) > \frac{1}{1 + S_1}.\]

{\bf Encoding at the relay and decoding at the destination:} the relay sends $X_2  = ( \widehat{T} + U_2 )\mod \Lambda$ and the destination receives 
\[ Y_3 = X_2 + S + Z_3, \]
and uses an MMSE estimator to decode $t$ by computing
\begin{align*}
Y_3' &= (  \alpha_2 Y_3 + U_q - U_2 ) \mod \Lambda \\
&= ( t  - Q_{\Lambda_q} (\alpha_2 S + U_q) + U_2 - (1-\alpha_2) X_2  \\
& \;\;\;\;\;\;\;\; + \alpha_2 S + \alpha_2 Z_3  + U_q - U_2 ) \mod \Lambda \\
&= ( t - (\alpha_2 S - U_q )\mod \Lambda_q - (1 - \alpha_2) X_2 + \alpha_2 Z_3 )\mod \Lambda \\
\end{align*}
where $(\alpha_2 S - U_q )\mod \Lambda_q$ is a random variable independent of all others which is uniformly distributed over $\mathcal{V}(\Lambda_q)$. Thus, $- (\alpha_2 S - U_q )\mod \Lambda_q$, $- (1 - \alpha_2) X_2$, and $\alpha_2 Z_3$ may be regarded as three independent noise terms with variances $\sigma^2 ( \Lambda_q)$,  $( 1 - \alpha_2)^2$ and $\alpha_2 ^2 \frac{1}{S_2}$, and approximated as Gaussian noise as in \cite{Erez:2004} when $n \rightarrow \infty$. 
In this last decoding step we see the effect of the interference``pre-cancellation'' at Node 1, as  illustrated in Fig.\ref{fig:model2}(b). In the Figure, the effect of the dithers is dropped for clarity and illustration purposes only, and is technically still required. 
 Choosing $\alpha_2 = \alpha_{2opt} = \frac{S_2}{S_2 + 1} $, the destination can decode $t$ when 
\begin{align}
R &< \frac{1}{2} \log \left( \frac{1}{ \frac{1}{ 1 + S_2} + \sigma^2(\Lambda_q) } \right)  \label{h} \\
&< \frac{1}{2} \log \left( \frac{1}{ \frac{1}{ 1 + S_2} + \frac{1}{ 1 + S_1} } \right) \label{g} \\
&= \frac{1}{2} \log \left( \frac{S_1S_2 + S_1 + S_2+1}{S_1 + S_2 + 2} \right). \label{i}
\end{align}
Observe that the constraint \eqref{con1} is always looser than the constraint \eqref{h}. 

\section{Conclusion}
\label{sec:conclusion}
We have proposed two nested-lattice code and DF-based achievability schemes for two two-hop channels with interference, where the interference is not known at the Tx or Rx of the link which experiences it. This renders the problem different from Costa's classical ``dirty-paper coding'' result in which interference is know non-causally at the Tx of the link over which it is experienced. The two channel models and their respective achievability schemes presented here amount to a form of distributed interference-cancellation. Both achievability schemes effectively rely on the structure of the underlying nested lattice codes to cancel an ``integer'' part of the interference, while treating the ``residual'' of this quantization as noise. We expect this technique to be of use in the development of coding theorems for larger networks, and in particular networks with cognition, or channel state information available at certain nodes.

\section*{Acknowledgment}
The work of Y. Song and  N. Devroye was partially supported by NSF under award 1053933.
The contents of this article are solely the responsibility of the authors and do not necessarily represent the official views of the NSF. 
\bibliographystyle{IEEEtran}
\bibliography{refs}
\end{document}